\documentclass[epj]{svjour}
\usepackage{longtable}
\usepackage{epsfig,graphicx}
\usepackage{lscape}

\usepackage{amsmath}

\begin{document}

\title{Investigation of rare nuclear decays with BaF$_2$ 
crystal scintillator contaminated by radium}

\author{
P.~Belli\inst{1}
\and R.~Bernabei\inst{1,2,}\thanks{e-mail: rita.bernabei@roma2.infn.it}
\and F.~Cappella\inst{3,4}
\and V.~Caracciolo\inst{5}
\and R.~Cerulli\inst{5}
\and F.A.~Danevich\inst{6}
\and A.~Di~Marco\inst{2}
\and A.~Incicchitti\inst{3}
\and D.V.~Poda\inst{6}
\and O.G.~Polischuk\inst{3,6}
\and V.I.~Tretyak\inst{3,6}
}

\institute{
INFN, Sezione di Roma ``Tor Vergata'', I-00133 Rome, Italy
\and Dipartimento di Fisica, Universit\`{a} di Roma ``Tor Vergata'', I-00133 Rome, Italy
\and INFN, Sezione di Roma, I-00185 Rome, Italy
\and Dipartimento di Fisica, Universit\`{a} di Roma ``La Sapienza'', I-00185 Rome, Italy
\and INFN, Laboratori Nazionali del Gran Sasso, I-67100 Assergi (AQ), Italy
\and Institute for Nuclear Research, MSP 03680 Kyiv, Ukraine
}

\date{Received: date / Revised version: date}

\abstract{
The radioactive contamination of a BaF$_2$ scintillation crystal
with mass of 1.714 kg was measured over 101 hours in the low-background DAMA/R\&D
set-up deep underground (3600 m w.e.) at the Gran Sasso National Laboratories of
INFN (LNGS, Italy). The half-life of $^{212}$Po (present in the crystal
scintillator due to contamination by radium) was measured as
$T_{1/2}(^{212}$Po) = 298.8$\pm$0.8(stat.)$\pm$1.4(syst.) ns 
by analysis of the events' pulse profiles.
The $^{222}$Rn nuclide is known as 100\% decaying via emission of $\alpha$ particle
with $T_{1/2}$ = 3.82 d; however,
its $\beta$ decay is also energetically allowed with $Q_\beta = 24\pm21$ keV. 
Search for decay chains of events
with specific pulse shapes characteristic for $\alpha$ or for $\beta/\gamma$ signals and with
known energies and time differences allowed us to set, for the first time, the limit
on the branching ratio of $^{222}$Rn relatively to $\beta$ decay as $B_\beta < 0.13$\% at 90\% C.L.
(equivalent to limit on partial half-life $T_{1/2}^\beta > 8.0$ y).
Half-life limits of $^{212}$Pb, $^{222}$Rn and $^{226}$Ra relatively to $2\beta$
decays are also improved in comparison with the earlier results. 
}


\maketitle
 
\section{Introduction}

The BaF$_2$ crystal is a promising scintillator for different applications, including 
detection of high energy gamma rays \cite{Rat02} and neutrons \cite{Gue09}.
The scintillation material is also widely used in medicine for positron emission
tomography (PET) \cite{Sel12}. The compound looks a promising detector to
search for double beta (2$\beta$) decay of barium. 2$\beta$ decay is a transformation of
nuclide $(A,Z)$ to $(A,Z\pm2)$ with simultaneous emission of two electrons or 
positrons and two (anti)neutrinos. This rare nuclear decay is allowed by the
Standard Model (SM). On the other hand, neutrinoless double $\beta$ decay is forbidden
by the SM due to violation of the lepton number $L$ by 2 units, but it is predicted 
by many SM extensions. It is considered as one of priority topics in current
nuclear and particle physics because it allows one to test 
the nature of neutrino (Dirac or Majorana particle),
the $L$ conservation, 
the absolute scale and the hierarchy of the neutrino masses, the existence of right-handed
admixtures in the weak interaction, the existence of Majorons and other interesting subjects
\cite{DBD-rev}. Natural barium contains two potentially 2$\beta$ active isotopes, 
$^{130}$Ba ($Q_{2\beta} = 2618.7(2.6)$ keV) and 
$^{132}$Ba ($Q_{2\beta} = 844.0(1.1)$ keV) \cite{Wan12,Tre95}. The
$^{130}$Ba isotope is of particular interest because of two reports on the
observation in geochemical experiments of double electron capture with half-life 
$T_{1/2} = (2.2\pm0.5)\times10^{21}$ y \cite{Mes01} and 
$T_{1/2} = (6.0\pm1.1)\times10^{20}$ y \cite{Puj09}.
The first direct laboratory search for $2\beta$ decays of $^{130}$Ba was performed
with BaF$_2$ crystal scintillator in Ref. \cite{Cer04}, where only $T_{1/2}$ limits
were obtained on the level of $\simeq10^{17}$ yr. Thus, further R\&D's for BaF$_2$
detectors are desirable.

Typically, high level of radioactive contamination of BaF$_2$ scintillation crystals
by uranium and thorium is the main source of background of the detectors
\cite{Cer04}; however, this feature allows us to use the detector for measurements
of some short-lived isotopes in U/Th chains (e.g. of $^{212}$Po). Results of
measurements of radioactive contamination of a large volume BaF$_2$ crystal
scintillator are presented in this work (section 2). We have also derived a half-life value of
$^{212}$Po from the data by using pulse-shape analysis of $^{212}$Bi -- $^{212}$Po
events (section 3).
In section 4, search for decay chains of events
with specific pulse shapes characteristic for $\alpha$ or for $\beta/\gamma$ signals,
and with known energies and time differences allowed us to set, for the first time, the limit
on the $\beta$ decay of $^{222}$Rn.
Half-life limits of $^{212}$Pb, $^{222}$Rn and $^{226}$Ra relatively to $2\beta$
decays are also improved in comparison with the earlier results \cite{Tre05}.

\section{Experimental measurements and data analysis}

\subsection{Experiment}

The radioactive contamination of a BaF$_2$ crystal ($\oslash 3''\times 3''$, 1.714 kg) was measured
over 101 hours in the low-background DAMA/R\&D set-up deep underground (3600 m
w.e.) at the Labotatori Nazionali del Gran Sasso of INFN (LNGS, Italy). The
BaF$_2$ crystal scintillator was viewed through two light-guides
($\oslash3''\times100$ mm) by two low radioactive $3''$ photomultipliers
(PMT, ETL 9302FLA). The detector was surrounded by Cu bricks and sealed in a low
radioactive air-tight Cu box continuously flushed by high purity nitrogen gas
to avoid the presence of residual
environmental radon. The Cu box was surrounded by a passive shield made of high
purity Cu, 10 cm of thickness, 15 cm of low radioactive lead, 1.5 mm of cadmium
and 4 to 10 cm of polyethylene/paraffin to reduce the external background. The
shield was contained inside a Plexiglas box, also continuously flushed by high
purity nitrogen gas. 

An event-by-event data acquisition system is operative in the set-up.
In details, the output signals of each PMT, after being pre-amplified,
were summed and sent to a 1 GSample/s Transient Digitizer (TD, Acqiris DC2 model) 
that recorded the signal profile over a time window of 4000 ns.
The preamplifier has 0 -- 250 MHz bandwidth, a factor 10 gain and a voltage
integral linearity 0.2\%. Therefore no sizable distortion of the PMT signals is expected.
A leading edge discriminator with threshold about --25 mV provided discriminated signal
for each PMT. The feed of the discriminator was obtained after filtering the PMT signal
by an Ortec Time Filter Amplifier (TFA). The trigger of the acquisition and of the
TD was obtained by requiring the coincidence of the 2 PMT discriminated signals in a 50 ns time window.
In such a way a hardware energy threshold of about 30 keV was obtained.
The trigger rate was around 75 counts/s. For each recorded event the area (``amplitude'')
of the pulse profile in a window of 1600 ns was calculated.

The energy scale of the BaF$_2$ detector and its energy resolution in the
range of interest have been determined by means of $^{22}$Na ($\gamma$ lines at 511, 1275 keV),
$^{137}$Cs (662 keV), $^{241}$Am (60 keV), $^{60}$Co (1173, 1333 keV), $^{133}$Ba
(356 keV) and $^{228}$Th (239, 2615 keV) sources. The energy resolution
(full width at half maximum, FWHM) for 662 keV $\gamma$ quanta of $^{137}$Cs
was 15.5\%, while for 511 and 1275 keV $\gamma$ lines of $^{22}$Na
source the energy resolution was 16.4\% and 10.8\%, respectively (see Fig.~1\footnote{In
all the figures the energy is given in $\gamma$ scale.}). The energy dependence of the energy
resolution can be approximated as: 
FWHM$_\gamma$ (keV) $= [397(54) + 15.6(3)\times E_\gamma]^{1/2}$, 
where $E_\gamma$ is the energy of the $\gamma$ quanta in keV.

\nopagebreak
\begin{figure}[htb]
\begin{center}
\mbox{\epsfig{figure=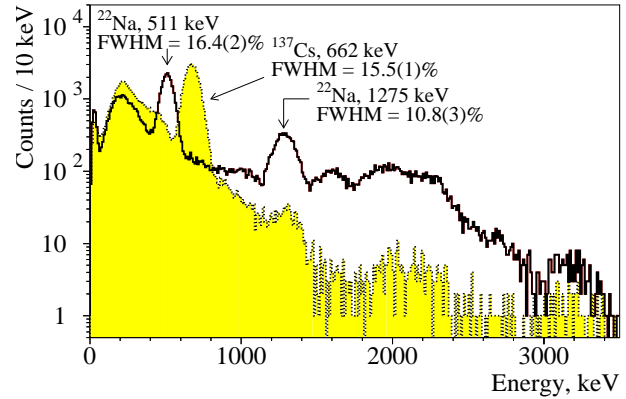,width=8.0cm}}
\caption{(Color online) Energy spectra accumulated with the BaF$_2$ detector
when irradiated by the $^{137}$Cs and $^{22}$Na $\gamma$ sources.
The peak at the energy of $\simeq1.3$ MeV in the $^{137}$Cs spectrum is due to
pile-ups of events. The peculiarities above $\simeq1.3$ MeV are caused by contamination
of the crystal by Ra.}
\end{center}
\end{figure}

We would like to make here a remark on equivalence of the energy scale for $\gamma$ quanta
and $\beta$ particles. Some scintillators, e.g. the liquid organic ones,
have quite big non-proportionality of the relative light output $L(E)/E$.
The examples can be seen in \cite{Bac02} for the Borexino liquid scintillator (LS)
and in \cite{Abe11} for the Double Chooz LS. 
Value of $L(E)/E$ is $90-95$\% at $\simeq100$ keV in comparison with 100\% 
at $\simeq1000$ keV and quickly drops further at lower energies. Liquid organic scintillators consist 
mainly from light elements C and H with small cross-sections of interaction with $\gamma$'s, 
and -- as an example -- 1 MeV $\gamma$ quantum has a few Compton scatterings before its total 
absorption in LS. Each of the produced Compton electrons has only part of the total 
energy of 1 MeV, and thus lower $L(E)/E$ value. When scintillation signals from all these 
electrons are collected, total signal in result has a smaller amplitude in comparison 
with that is expected from single 1 MeV electron. Thus, in big-size scintillators non-proportionality 
leads to worse energy resolution and to energy shift for $\gamma$ quanta. For example, 
in \cite{Bac03} position of the 1461 keV $^{40}$K 
$\gamma$ peak corresponds to 1360 keV of energy deposited for an electron. Size of 
scintillator is important because in smaller detectors $\gamma$ quantum will give 
the peak of total absorption preferably through photoeffect producing only one 
electron with high energy and $L(E)/E$ value closer to 100\%. 

For BaF$_2$ scintillator, non-proportionality in $L(E)/E$ is much lower than 
that for organic LS, as demonstrated e.g. in \cite{Dor95,Kho12}.
In addition, volume of our BaF$_2$ crystal is $3-4$ orders of magnitude lower than 
those of the Borexino and Double Chooz detectors. 
Thus, one can expect much lower difference between $\gamma$ 
and $\beta$ signals in the BaF$_2$ (which additionally is masked by rather poor 
energy resolution). This effect surely exists in our 
BaF$_2$ detector, but in the present article it is not taken into account 
and considered as not very important for the aims of this work.

\subsection{Pulse-shape discrimination of $\alpha$, $\beta/\gamma$ and Bi-Po events}

Scintillation signals from events of different origin ($\alpha$ particles;
$\gamma$ quanta or $\beta$ particles; PMT noise; etc.) have different
time profiles, and this can be used for their discrimination.
We utilize here modification of a pulse-shape discrimination (PSD) technique based on 
a mean time of events (see e.g. \cite{Cer04,Pec99}). 
Time profile of an event, stored in 4000 channels with 1 ns channel's width, is used to
calculate its mean time as:
$\langle t \rangle = \sum a_i \cdot t_i / \sum a_i$,
where the sum is over time channels $i$ starting from the origin of the pulse up 
to certain time, and $a_i$ is the digitized amplitude (at the time $t_i$) of a given signal.

The scatter plot of the mean time versus energy for background runs is shown in 
Fig.~2 which demonstrates pulse-shape discrimination ability of the BaF$_2$ detector.
Mean time distributions (similar to the one presented in Inset of Fig.~2) were 
built for 18 energy intervals of 100 keV wide in the range of $1200-3000$ keV 
using the data of the background measurements. 
Then the mean time distributions were fitted by two Gaussians 
which represent $\alpha$'s and $\gamma$'s, with the centre and width dependent on 
energy.
The following dependencies were obtained: \\
$\tau_\gamma(E) = 556(12)$, \\
$\sigma_\gamma(E) = 0.9(1)+653(23)/\sqrt{E}$, \\
$\tau_\alpha(E) = 490(13)+6(3)\times10^{-3} \times E$, \\
$\sigma_\alpha(E) = 3.1(1)+512(22)/\sqrt{E}$, \\
where $E$ is in keV, and $\tau$ and $\sigma$ are in ns.
Using these dependencies, $\pm3\sigma$ contours were calculated where 99\% of the 
corresponding events were contained (see Fig.~2).

\nopagebreak
\begin{figure}[htb]
\begin{center}
\mbox{\epsfig{figure=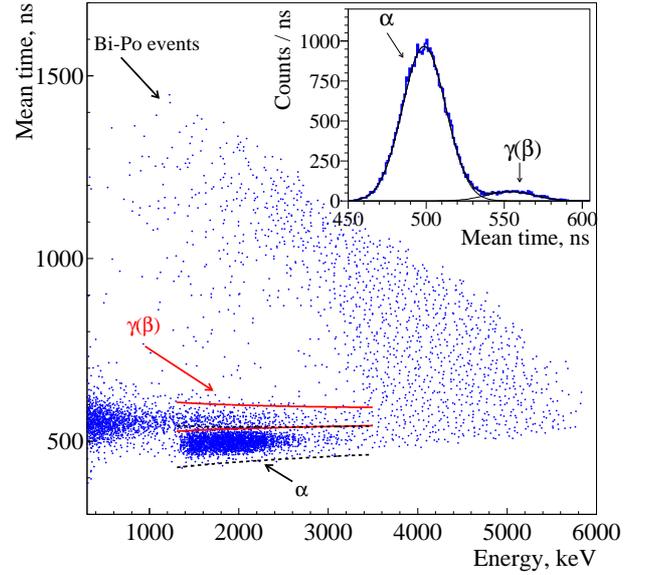,width=8.0cm}}
\caption{(Color online) Scatter plot of the mean time versus energy
accumulated by the BaF$_2$ scintillation detector during one background run ($\simeq2$ h).
The contours give regions where 99\% of $\alpha$ or $\beta/\gamma$ events are
concentrated.
The population of events in the energy interval $\simeq 1-6$ MeV with mean time values mainly 
above the $\beta/\gamma$ and $\alpha$ regions are caused by the decays of the fast 
$^{212}$Bi--$^{212}$Po sub-chain of $^{228}$Th (Bi-Po events, see text). 
(Inset) The mean time spectrum in the energy interval 2000 -- 2100 keV. The $\alpha$ and
$\beta/\gamma$ events distributions are fitted by Gaussian functions
(solid lines).}
\end{center}
\end{figure}

If the time interval between two subsequent signals is shorter than 4000 ns, the data acquisition system
will record them as one event. Nevertheless, such events can be recognized by analysis of their
time profile. 
Events in the fast chains $^{212}$Bi--$^{212}$Po (from $^{232}$Th family), with 
$T_{1/2}(^{212}$Po) = 299 ns and 
$^{214}$Bi--$^{214}$Po (from $^{238}$U), with
$T_{1/2}(^{214}$Po) = 164.3 $\mu$s belong to this category (the so-called Bi-Po events).
An example of the $^{212}$Bi $\rightarrow$ $^{212}$Po $\rightarrow$ $^{208}$Pb event in the
BaF$_2$ scintillator is presented in Fig.~3.
The Bi-Po events can be separated from $\alpha$ and $\beta/\gamma$ events by the pulse-shape 
discrimination because of their specific time profile which leads to different mean time distribution
(see Fig.~2). 
The time intervals between $\beta$ events (of
$^{212}$Bi or $^{214}$Bi) and subsequent $\alpha$ events (of $^{212}$Po or
$^{214}$Po) were obtained by analysis of the pulse profiles. 

\nopagebreak
\begin{figure}[htb]
\begin{center}
\mbox{\epsfig{figure=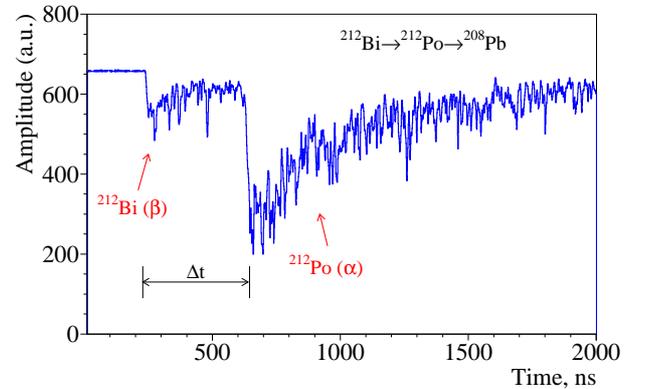,width=8.0cm}}
\caption{(Color online) Example of the Bi-Po event in the BaF$_2$ scintillator.}
\end{center}
\end{figure}

The background energy spectrum accumulated over 101 hours with the BaF$_2$
crystal scintillator is presented in Fig.~4. A substantial increase of the
counting rate in the energy interval 1.3 -- 3.4 MeV is due to $\alpha$ activity
of $^{238}$U and $^{232}$Th daughters from radium contamination (see below) of
the BaF$_2$ crystal.
Separation of the $\alpha$ and $\beta/\gamma$ signals was done event by event 
using the PSD technique (corresponding spectra are also shown in Fig.~4).

\nopagebreak
\begin{figure}[htb]
\begin{center}
\mbox{\epsfig{figure=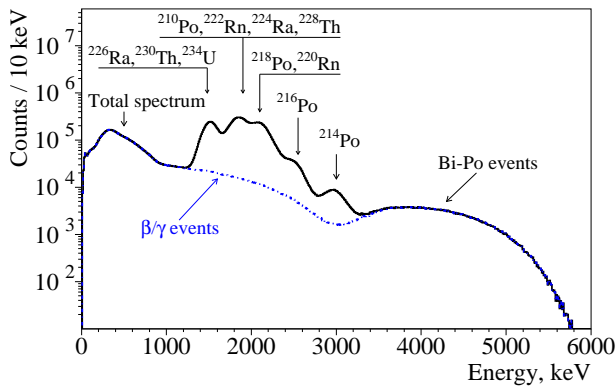,width=8.0cm}}
\caption{(Color online) Background energy spectrum of the BaF$_2$
scintillator collected during 101 hours. The spectrum of $\beta$/$\gamma$
events separated by the pulse-shape discrimination is shown by dotted line.}
\end{center}
\end{figure}

The energy spectrum of $\alpha$ events selected from the background data with the
help of the pulse-shape discrimination is presented in Fig.~5. 
We use asymmetric cut which selects 98\% of $\alpha$ 
events\footnote{We have selected events in the interval of the 
mean time values from 300 ns up to the value of $\tau_\alpha$ where 98\% 
of the integral is reached (taking into account the $\tau_\alpha(E)$ and 
$\sigma_\alpha(E)$ energy dependencies).}; 
around 18\% of
$\gamma/\beta$ events also are present in the resulting data.
The $\alpha$ spectrum was
fitted by a model built by $\alpha$ peaks from $^{235}$U, $^{238}$U and $^{232}$Th and their
daughters, assuming broken equilibrium in the chains. 
The equilibrium can be broken due to different chemical properties of
nuclides in U/Th chains (in comparison with those of barium) and 
relatively big half-lives of some nuclides in the chains.
Therefore the activities of
the following nuclides and sub-chains ($^{232}$Th, $^{228}$Th--$^{212}$Pb;
$^{238}$U, $^{234}$U, $^{230}$Th, $^{226}$Ra--$^{214}$Po, $^{210}$Po; $^{235}$U,
$^{231}$Pa, $^{227}$Ac--$^{211}$Bi) were taken as free parameters of the fit.
Furthermore, values of the energy resolution of the detector to $\alpha$ particles and
$\alpha/\beta$ ratio (relative light output for $\alpha$ particles as compared with that for
$\beta$ particles ($\gamma$ rays) of the same energy) 
were also introduced into the fit as free parameters.
According to the fit, the $\alpha/\beta$ ratio for the BaF$_2$ scintillation detector
depends on energy of $\alpha$ particles as 
$\alpha/\beta = 0.200(1) + 0.0245(1)\times E_\alpha$, 
where $E_\alpha$ is energy of $\alpha$ particles in MeV.
The energy resolution for $\alpha$ particles is 
FWHM$_\alpha$ (keV) = $(28 \times E^\gamma_\alpha)^{1/2}$,
where $E^\gamma_\alpha$ is the energy of $\alpha$ particles in $\gamma$ scale (in keV).
Some difference between the fit and the spectrum of the selected $\alpha$'s in Fig.~5 
can be due to not perfectly Gaussian shape of the $\alpha$ peaks, nonlinear 
dependence of the $\alpha/\beta$ ratio on energy in wide energy
interval, some broadening and shift of 
the $^{222}$Rn and $^{218}$Po peaks due to nonuniformity of light collection 
caused by diffusion of radon in the crystal.
The radioactive contaminations of the BaF$_2$ crystal obtained from the fit are
presented in Table~1. 

One can conclude that the BaF$_2$ crystal is contaminated by radium 
which was not removed from the BaF$_2$ during the material preparation and the
crystal growth because of chemical similarity of barium and radium,
while all other elements which belong to U/Th chains were effectively
removed that resulted in broken secular equilibrium. 
Residual contamination by $^{226}$Ra ($T_{1/2}=1600$ y)
lead to related activities of daughter $\alpha$
decaying $^{222}$Rn and $^{218}$Po (in equilibrium with $^{226}$Ra).
Area of $^{214}$Po $\alpha$ peak is smaller in comparison with the peak 
of $^{226}$Ra (see Fig.~5) due to the dead time of the detector of 1.65 ms, 
big compared with $^{214}$Po half-life (164 $\mu$s).
Equilibrium at $^{210}$Pb is also broken that is related with its big half-life (22.3 y) in 
comparison with the time passed from production of the BaF$_2$ crystal (few years);
this results also in lower activity of its $\alpha$ decaying daughter $^{210}$Po in Fig.~5.
Similarly, presence of $^{228}$Th, $^{224}$Ra, $^{220}$Rn, $^{216}$Po, $^{212}$Bi in the 
$\alpha$ spectrum can be explained by residual contamination of the BaF$_2$ crystal
by $^{228}$Ra ($T_{1/2}=5.75$ y).
Contribution from $^{212}$Bi in Fig.~5 is smaller because of its 
$\simeq36\%$ branching in the $^{228}$Ra chain. $^{228}$Ra itself does not contribute
to the $\alpha$ spectrum because it decays with emission of $\beta$'s and $\gamma$'s.

\nopagebreak
\begin{figure}[htb]
\begin{center}
\mbox{\epsfig{figure=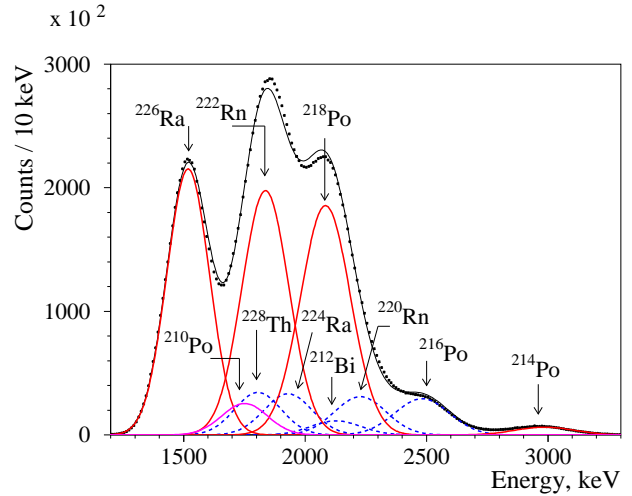,width=8.0cm}}
\caption{(Color online) Energy spectrum of $\alpha$ events selected by 
pulse-shape discrimination from the data of low-background measurements with the
BaF$_2$ crystal over 101 hours (points). Fit of the data by the model built
from $\alpha$ decays of $^{238}$U and $^{232}$Th daughters is shown by solid black line
(individual components of the fit are presented by colored lines).}
\end{center}
\end{figure}

\begin{table}[htb]
\caption{Radioactive contaminations of the BaF$_2$ crystal.
Limits are given at 90\% C.L.}
\begin{center}
\begin{tabular}{lll}
\hline
Chain      & Nuclide    & Activity,  \\
~          & ~          & Bq/kg      \\
\hline
~ & ~ & ~ \\
$^{232}$Th & $^{232}$Th & $<$0.004   \\
~          & $^{228}$Th & 1.35(6)    \\
~ & ~ & ~ \\
$^{238}$U  & $^{238}$U  & $<$0.0002  \\
~          & $^{226}$Ra & 7.8(1)     \\
~          & $^{210}$Pb & 0.99(1)    \\
~ & ~ & ~ \\
$^{235}$U  & $^{235}$U  & $<$0.0006  \\
~          & $^{231}$Pa & $<$0.0007  \\
~          & $^{227}$Ac & $<$0.07    \\
~ & ~ & ~ \\
\hline
\end{tabular}
\end{center}
\end{table}

\subsection{Analysis of Bi-Po events}

The Bi-Po events were selected as events with energy $E > 30$ keV and the
time difference between the first and second signals in the range of 30 ns to 3000 ns;
discrimination by mean time method was not used.

The energy spectra of the first events ($\beta$ decay of $^{212}$Bi with
$Q_\beta = 2254$ keV and $^{214}$Bi with $Q_\beta = 3272$ keV) and
of the second events ($\alpha$ decay of $^{212}$Po with $Q_\alpha = 8954$ keV
and $^{214}$Po with $Q_\alpha = 7833$ keV) selected from the Bi-Po pulse
profiles are presented in Fig.~6.

\nopagebreak
\begin{figure}[htb]
\begin{center}
\mbox{\epsfig{figure=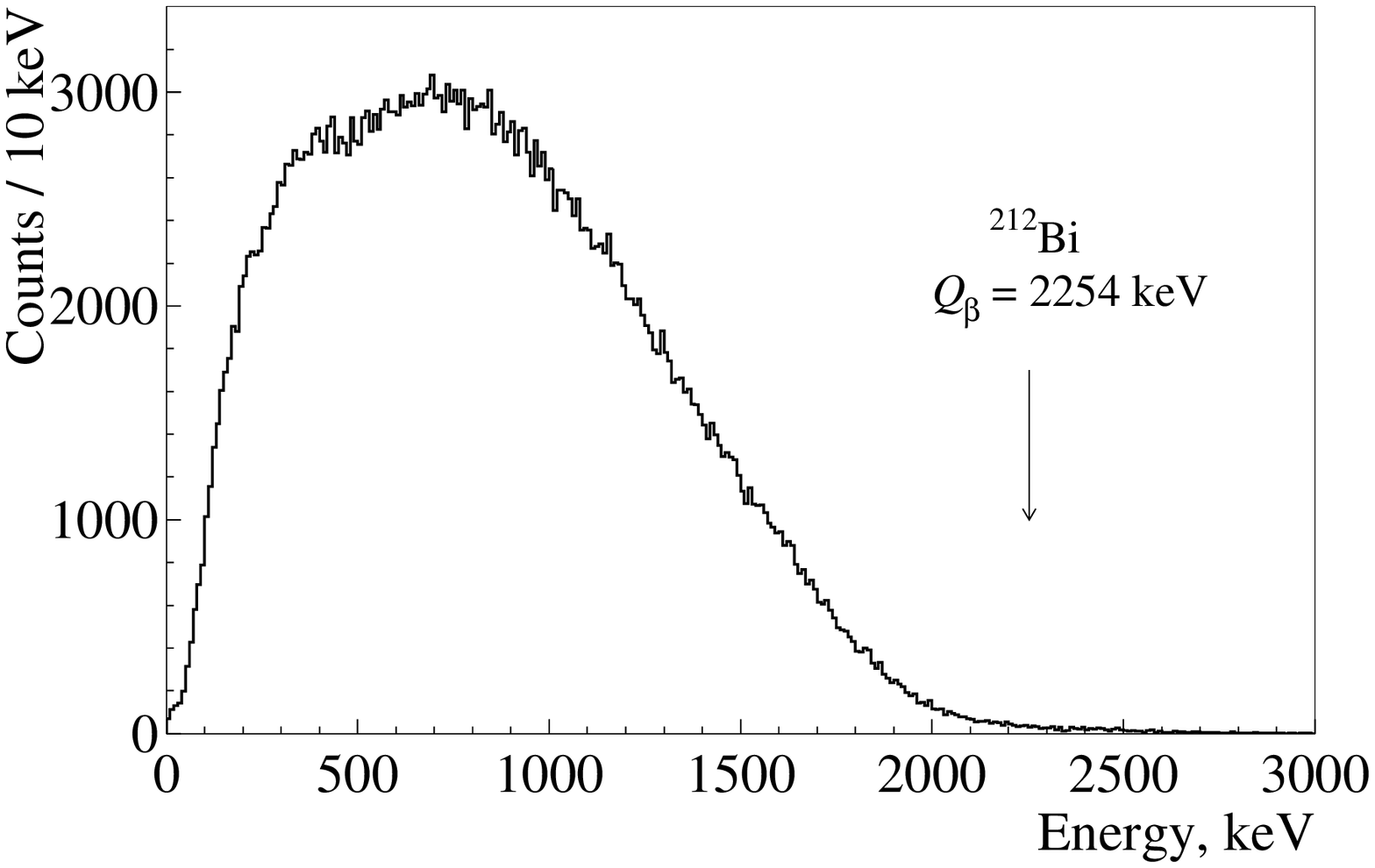,width=8.0cm}} \\
\mbox{\epsfig{figure=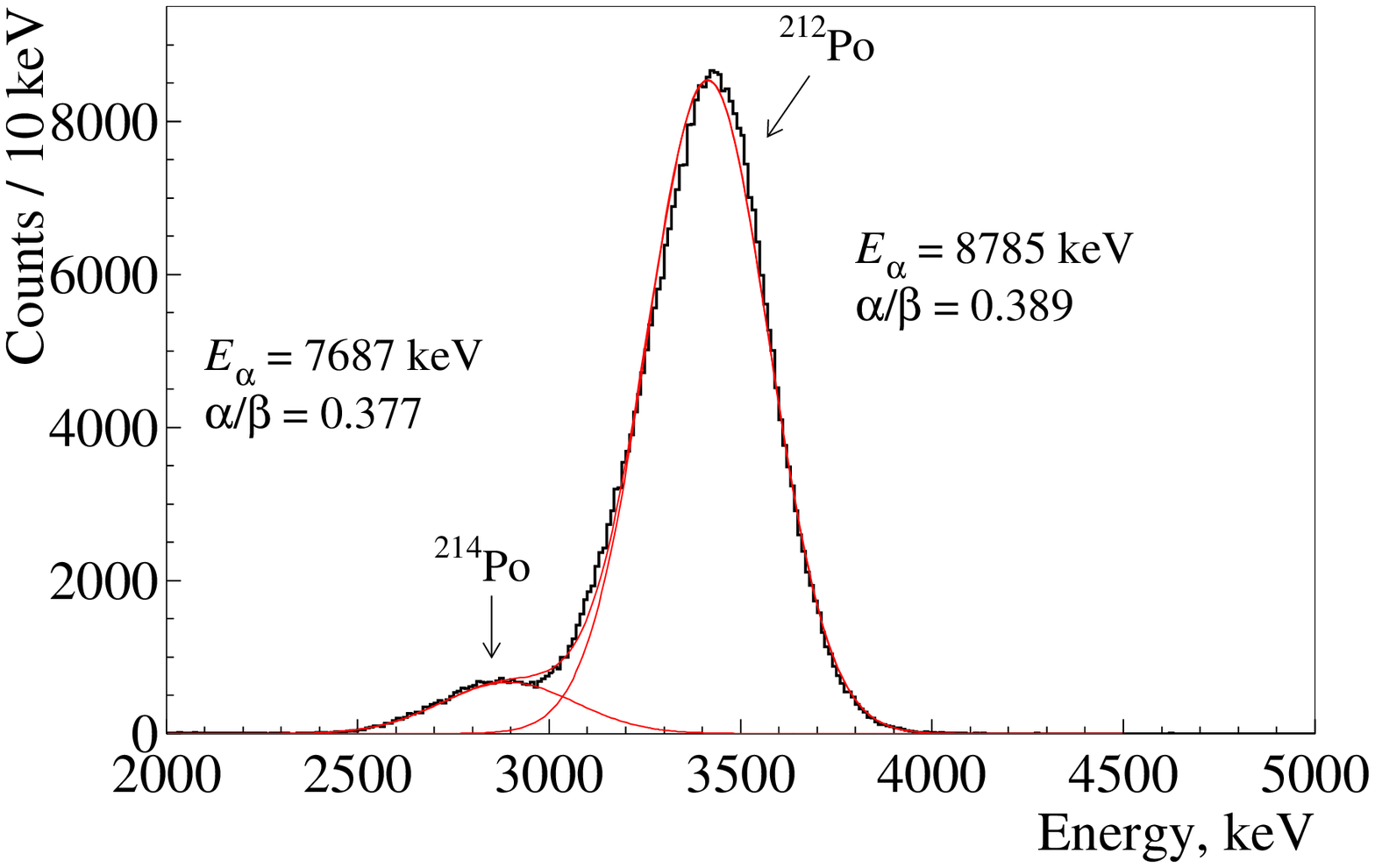,width=8.0cm}}
\caption{(Color online) Energy spectra of the first (mainly
$^{212}$Bi with $Q_\beta = 2254$ keV; top) and of the second
(mainly $^{212}$Po, $Q_\alpha = 8954$ keV; bottom) events selected by the
pulse-shape analysis of the background data accumulated with the BaF$_2$
crystal scintillator over 101 hours.
Red lines show individual components of the fit, while black line represents 
the total fit.
Beta spectrum below $\simeq300$ keV is distorted by the energy threshold effects.}
\end{center}
\end{figure}

We have estimated activity of $^{228}$Th from the Bi-Po analysis as
1.17(12) Bq/kg, which is in reasonable agreement with the result obtained from
the fit of the alpha spectrum presented in Fig.~5.

\section{Half-life of $^{212}$Po}

The distribution of the time intervals between two subsequent signals in the 
Bi-Po events is presented in Fig.~7. In order to suppress the contribution of the
$^{214}$Bi--$^{214}$Po chain (with the half-life of $^{214}$Po 
$T_{1/2} = 164.3~\mu$s), the energy of the second event was selected in the
energy interval 3000 -- 3800 keV (see bottom part in Fig.~6). 
In the following, an energy threshold of
300 keV was chosen for the first events to decrease jitter of the event time
determination. The time spectrum was fitted by sum of two exponential functions
that represent the decays of $^{212}$Po and $^{214}$Po and constant 
which describes contribution from randomly coincident events
(with the total counting rate of 75 counts/s, this contribution is $<1\%$).
The half-life of the second exponent was not exactly fixed but restricted in 
the interval of $T_{1/2} = 164.3\pm2.0~\mu$s taking into account the table 
uncertainty of the $^{214}$Po $T_{1/2}$ \cite{ToI98}.

To estimate a systematic error of the half-life value, the time 
spectra (with time bins of 1, 2 and 3 ns per channel) were fitted by the 
chi-square method in 30 different time intervals (for each binning).
The starting point in the fit varied from 100 ns to 250 ns, and the end point 
from 1350 ns to 1550 ns.
All the fits gave the $\chi^2$/n.d.f. values in the range of
0.92 -- 1.15 (where n.d.f. is the number of degrees of freedom).
The obtained values of $^{212}$Po half-life lay in quite narrow
range of 297.4 -- 299.8 ns with an average value of 298.8$\pm$0.8(stat.) ns. 
Different binning of the time spectrum contributes to the systematic error for 1.05 ns,
and the change of the fitting interval provides 0.96 ns; summing them quadratically, they
give the systematic error of 1.4 ns. 
This value is considered as reliable: for example, the fitting procedure
with the time bin increased to 10 ns gives an average $T_{1/2}$ value of 
300.2$\pm$0.8(stat.) ns which is inside the systematic error. 

The obtained final half-life 
$T_{1/2}$ = 298.8 $\pm$0.8(stat.) $\pm$1.4(syst.) ns is in an
excellent agreement with the table value $299\pm2$ ns \cite{ToI98,Bro05}, and in
reasonable agreement with the recent result of the Borexino collaboration 
$T_{1/2}$ = 294.7 $\pm$0.6(stat.) $\pm$0.8(syst.) ns \cite{Bel13}.

\nopagebreak
\begin{figure}[htb]
\begin{center}
\mbox{\epsfig{figure=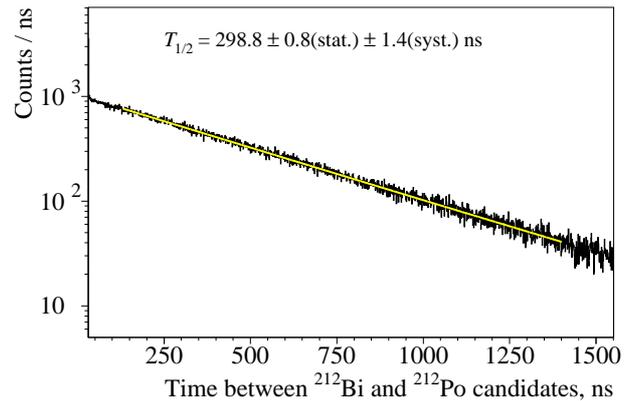,width=8.0cm}}
\caption{(Color online) The time distribution for the fast sequence of
$\beta$ ($^{212}$Bi) and $\alpha$ ($^{212}$Po) decays selected
from the data accumulated with the BaF$_2$
scintillation detector over 101 hours. The obtained half-life
$T_{1/2}$ = 298.8$\pm$0.8(stat.)$\pm$1.4(syst.) ns is in good
agreement with the table value $299\pm2$ ns \cite{ToI98,Bro05}.}
\end{center}
\end{figure}

\section{Search for rare $\beta$ and $2\beta$ decays in U/Th chains}

The quite-high level of contamination of the BaF$_2$ detector by Ra gives the possibility
to search for rare $\beta$ and $2\beta$ decays of some nuclides in U/Th chains, which 
cannot be easily studied in the usual way since they  
fastly decay through $\alpha$ and/or $\beta$ channels and cannot be accumulated in
big quantities. The idea to search for $2\beta$ decay of unstable nuclides was discussed
in \cite{Tre05}, motivated by the fact that sometimes unstable nuclides can have quite
big energy release $Q_{2\beta}$ ($\simeq43$ MeV for $^{19}$B and $^{22}$C) while for
the ``standard'' 69 double $\beta$ nuclides (present in natural mixture of elements) the maximal
value is $\simeq4.3$ MeV ($^{48}$Ca) \cite{Tre95}. The probability of neutrinoless ($0\nu$)
double $\beta$ decay for big energy releases
depends on $Q_{2\beta}$ value roughly as $Q_{2\beta}^5$, and 
the probability of two neutrino ($2\nu$) double $\beta$ decay as $Q_{2\beta}^{11}$.
Thus these processes proceed faster for nuclides with high $Q_{2\beta}$. 
While no nuclide for immediate breakthrough was found in \cite{Tre05}, few interesting
candidates were identified ($^{42}$Ar, $^{126}$Sn, $^{208}$Po), and first $T_{1/2}^{2\beta}$ limits
were set for nuclides in U/Th chains by analysis of energy spectra of CaWO$_4$, CdWO$_4$ and
Gd$_2$SiO$_5$ crystal scintillators. 

The half-life limit on $2\beta$ decay can be calculated with the following formula:
$\lim T_{1/2}^{2\beta} = \varepsilon \cdot \ln 2 \cdot N \cdot t/\lim S,$ 
where $\varepsilon $ is the efficiency to detect the 2$\beta $ process,
$N$ is the number of the $2\beta$ decaying nuclei,
$t$ is the time of measurements, and $\lim S$ is the number of
2$\beta $ events which can be excluded with a given
confidence level on the basis of experimental data.
The number of unstable nuclei in radioactive chain can be determined 
from its decay rate (which is the same for all nuclides in equilibrium):
$R^{\alpha/\beta} = dN/dt = \ln 2 \cdot N / T_{1/2}^{\alpha/\beta}$.
Here $T_{1/2}^{\alpha/\beta}$ is the isotope's half-life for
the usual $\alpha$ or $\beta$ decay (the small contribution from
2$\beta $ decay can be neglected). Finally, one can get:

\begin{equation}
\lim T_{1/2}^{2\beta} = \varepsilon \cdot t \cdot R^{\alpha/\beta} \cdot 
T_{1/2}^{\alpha/\beta} / \lim S.
\end{equation}

In the present work we improve $T_{1/2}^{2\beta}$ limits for some nuclides by
searching for chains of decay resulting from $2\beta$ decay of initial nuclide.
Moreover, we give the first limit for the single $\beta$ decay of $^{222}$Rn.

\subsection{First search for $\beta$ decay of $^{222}$Rn}

While $^{222}$Rn is considered as 100\% decaying through emission of $\alpha$ particle \cite{ToI98,Sin11}, 
its single beta decay is also energetically allowed, with the energy release
$Q_\beta=24\pm21$ keV in accordance with the last atomic mass tables \cite{Wan12}.
The ground state to the ground state $\beta$ decay $^{222}$Rn(0$^+$) $\to$ $^{222}$Fr(2$^-$)
proceeds with change in spin and parity $\Delta J^{\Delta \pi}=2^-$, and thus is 
classified as first forbidden unique. The expected half-life can be estimated in the 
following way. 

A recent compilation of Log~$ft$ values \cite{Sin98} gives the average value Log~$ft=9.5\pm0.8$
for all known 216 first forbidden unique $\beta$ decays. 
Using the LOGFT tool at the National Nuclear Data Center, USA \cite{logft}, one can find that
the central value of Log~$ft=9.5$ corresponds to half-life 
$T_{1/2}^\beta=4.8\times10^5$ y for $Q_\beta=24$ keV. Taking into account the uncertainties in
the $Q_\beta$ value, the half-life is equal to
$6.7\times10^4$ y for $Q_\beta=45$ keV and
$2.4\times10^8$ y for $Q_\beta=3$ keV. 

To our knowledge, there were no previous attempts to experimentally search for $\beta$ decay 
of $^{222}$Rn. This is related, in particular, with the small half-life relatively to $\alpha$ decay
(3.82 d \cite{Sin11}), the low $Q_\beta$ value (typically below the energy threshold) and 
the usually high background at low energies. 
To realize such a search with the help of the BaF$_2$ detector, we note that
$\beta$ decay of $^{222}$Rn leads to a chain of subsequent decays:
\begin{equation}\begin{split}
& ^{222}_{~86}\text{Rn}~\xrightarrow[24~\text{keV}]{\beta?~~\simeq4.8\times10^5~\text{y}}~
  ^{222}_{~87}\text{Fr}~\xrightarrow[2028~\text{keV}]{\beta~~14.2~\text{m}}~
  ^{222}_{~88}\text{Ra}~\xrightarrow[6679~\text{keV}]{\alpha~~38.0~\text{s}}~ \\
& ^{218}_{~86}\text{Rn}~\xrightarrow[7263~\text{keV}]{\alpha~~35~\text{ms}}~ 
  ^{214}_{~84}\text{Po}~\xrightarrow[7833~\text{keV}]{\alpha~~164.3~\mu\text{s}}~
  ^{210}_{~82}\text{Pb (22.3 y)}.
\end{split}\end{equation}

The quite-high activity of $^{226}$Ra (parent of $^{222}$Rn) of 13.4 Bq in the BaF$_2$ crystal, 
the difference in scintillation responses, which allows to discriminate $\alpha$ and $\beta/\gamma$ 
events by pulse-shape analysis,
and the knowledge of the expected energies and time differences between events
give the possibility to search for this chain in the accumulated data.
It should be noted, however, that:
(a) due to high total events rate in the BaF$_2$ detector (75 counts/s) it is unfeasible to look
for events with $T_{1/2}=14.2$ m ($^{222}$Fr $\beta$ decay) because of many events of 
other origin; at most, we should restrict ourselves by $\alpha$ decay of $^{222}$Ra with 
$T_{1/2}=38.0$ s; 
(b) unfortunately, we cannot use also $\alpha$ decay of $^{214}$Po with $T_{1/2}=164.3~\mu$s due
to the big dead time in the present measurements (1.65 ms), that results in a very low efficiency
(the probability of a $^{214}$Po decay after the dead time is only $9.1\times10^{-4}$) and in a low
sensitivity.
Practically, we can only use the chain of decays: $^{222}$Fr $\to$ $^{222}$Ra $\to$ $^{218}$Rn 
searching for the following sequence of events:

(1) an event with energy from 30 to 2207 keV ($^{222}$Fr $Q_\beta$ + FWHM$_\gamma$) and mean time 
which is characteristic for $\beta$ events (it should be in interval where 99\% of $\beta$ events
are expected);

(2) next event with energy between 2109 and 2623 keV ($^{222}$Ra $E_\alpha$ in the $\gamma$ 
scale of the BaF$_2$ scintillator $\pm$ FWHM$_\alpha$), 
mean time characteristic for $\alpha$ events and in the time interval 
[1.65 ms, 1.65 ms + $5\times38.0$ s];

(3) the last event with energy between 2398 and 2946 keV ($^{218}$Rn $E_\alpha^\gamma$ $\pm$ FWHM$_\alpha$), 
mean time characteristic for $\alpha$ events and in the time interval 
[1.65 ms, 1.65 ms + $5\times35$ ms].

There are $7.0\times10^5$ events in the accumulated data which satisfy all the above listed 
criteria simultaneously. The energy spectrum of the last events in the chain is shown in Fig. 8, top
(where also data in wider energy region $2100-3300$ keV are presented). 
It is obvious that it also contains additional events of other origin due to high counting rate
in the BaF$_2$. The maximal possible number of $^{218}$Rn $\alpha$ decays can be calculated very
conservatively just requiring that the expected gaussian distribution should not exceed
the experimental energy spectrum at any region, but in some proper interval, where theoretical
effect is closest to the experimental spectrum, their areas should be equal.
Such a curve is also shown in Fig. 8, top; the corresponding area is $1.3\times10^5$ counts
(which takes into account also uncertainty in the experimental statistics at 90\% C.L.).

\nopagebreak
\begin{figure}[htb]
\begin{center}
\mbox{\epsfig{figure=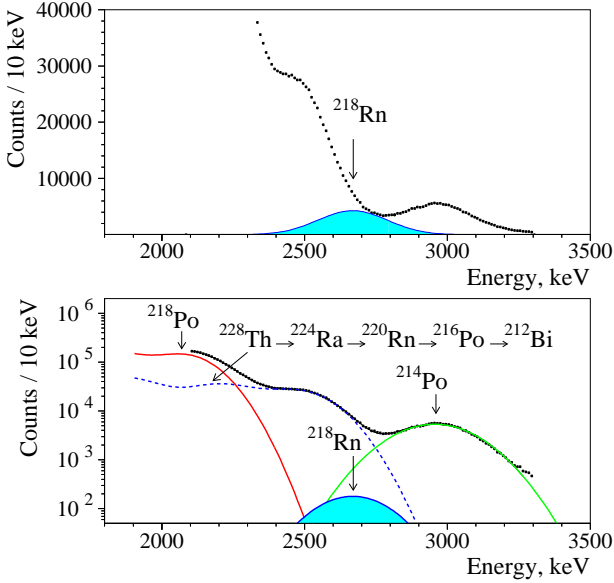,width=8.0cm}}
\caption{(Color online) Top: Energy spectrum of possible events of $^{218}$Rn $\alpha$
decay in the chain $^{222}$Fr $\to$ $^{222}$Ra $\to$ $^{218}$Rn (see text): 
experimental data and maximal effect consistent with the data.
Bottom: Fit of the selected data by sum of the background model built from 
$\alpha$ peaks of isotopes in U/Th chain together with the excluded $^{218}$Rn $\alpha$ peak.
}
\end{center}
\end{figure}

One can calculate limit on $T_{1/2}^\beta$ of $^{222}$Rn, using the formula 
similar to Eq. (1):

$$\lim T_{1/2}^\beta = \varepsilon \cdot t \cdot R^\alpha \cdot T_{1/2}^\alpha / \lim S,$$

\noindent and values 
$\lim S = 1.3\times10^5$,
$\varepsilon = 0.849$ (taking into account the time and energy intervals used and 
efficiency for pulse-shape discrimination),
$t = 101$ h,
$R^\alpha = 13.4$ Bq (we can use activity for $^{222}$Rn the same as for 
$^{226}$Ra because $^{222}$Rn is in equilibrium with $^{226}$Ra),
$T_{1/2}^\alpha = 3.8235$ d for $^{222}$Rn. The result is equal:
$T_{1/2}^\beta(^{222}$Rn$) > 122$ d at 90\% C.L.

However, it is evident that this estimation is too conservative: there is
no reason to ascribe all the events at $\simeq2.75$ MeV only to the effect searched for.
One can estimate the effect in a more realistic way, fitting
the spectrum by the sum of (1) some model which represents the background and (2)
gaussian with known center and width which corresponds to the $^{218}$Rn peak searched for.
In fact, shape of the spectrum of the selected possible $^{218}$Rn $\alpha$ events in 
Fig.~8 is similar to the spectrum of all $\alpha$ events shown in Fig.~5. 
The isotopes in $^{232}$Th and $^{238}$U chains randomly give contributions to the data 
of Fig.~8 due to big time interval used in the selection ($\simeq5\times38$ s) and high 
counting rate (75 counts/s). Thus, we build the background model from $\alpha$ peaks of 
nuclei in U/Th chains similarly to Fig.~5. Fit of the data in the energy interval
2110 -- 3260 keV (see Fig.~8, bottom) by this model and the $^{218}$Rn gaussian
results in the area of the gaussian $S = 3023 \pm 1476$ counts\footnote{The fit gives 
activities of $^{228}$Th and $^{226}$Ra as 1.5(1) Bq/kg 
and 7.6(2) Bq/kg, respectively, in good agreement with the results obtained by 
the fit of the total $\alpha$ spectrum (see Fig. 5 and Table 1).}. 
In accordance with the Feldman-Cousins procedure 
\cite{Fel98}, the limit on the area is $\lim S = 5444$ counts at 90\% C.L.
This gives the following $T_{1/2}^\beta(^{222}$Rn):

$$T_{1/2}^\beta(^{222}\text{Rn}) > 8.0~\text{y at 90\% C.L.}$$

\noindent
which corresponds to limit on branching ratio $B_\beta < 0.13\%$.

\subsection{$2\beta$ decay of $^{222}$Rn}

In the chain presented by Eq. (2), $^{222}$Rn transforms to $^{222}$Ra by two subsequent
$\beta$ decays: 
$$
^{222}_{~86}\text{Rn}~\xrightarrow[24~\text{keV}]{\beta~~\simeq4.8\times10^5~\text{y}}~~
^{222}_{~87}\text{Fr}~\xrightarrow[2028~\text{keV}]{\beta~~14.2~\text{m}}~~
^{222}_{~88}\text{Ra}.
$$

However, in principle, it can transform to $^{222}$Ra in one step through $2\beta$ decay:
$$
^{222}_{~86}\text{Rn}~\xrightarrow[2052~\text{keV}]{2\beta}~~
^{222}_{~88}\text{Ra}
$$

\noindent (there is no law which forbids this, at least for two neutrino $2\beta$ decay
allowed by the Standard Model). The energy release is equal $Q_{2\beta} = 2052$ keV, 
only slightly higher than $Q_\beta = 2028$ keV in $\beta$ decay of $^{222}$Fr.
The chain of decays $^{222}$Rn $\to$ $^{222}$Ra $\to$ $^{218}$Rn is very similar to the 
chain $^{222}$Fr $\to$ $^{222}$Ra $\to$ $^{218}$Rn, which we searched for in the previous 
section. 
The only difference is that we should search for the first event which has mean time 
characteristic for $\beta$ events (as previously) but its energy is from 0 to 2231 keV for
$2\beta2\nu$ decay of $^{222}$Rn or in the interval of $2052\pm180$ keV for its
$2\beta0\nu$ decay. Efficiency of selection of events is equal to 0.849 (0.841) for
$2\beta2\nu$ ($2\beta0\nu$) decay. 
The maximal number of events in the chain is found as $\lim S = 5.4\times10^3$
(both for $2\beta2\nu$ and $2\beta0\nu$ processes due to high counting rate in the BaF$_2$);
hence, the following limit can be achieved:

$$T_{1/2}^{2\beta(0\nu+2\nu)}(^{222}\text{Rn}) > 8.0~\text{y at 90\% C.L.}$$

The obtained value is better than those obtained in \cite{Tre05}
(2.8 y for $2\beta0\nu$ and 40 d for $2\beta2\nu$ at 68\% C.L.).
In terms of the branching ratio, the limit is equal $B < 0.13$\%.

\subsection{$2\beta$ decay of $^{226}$Ra}

$^{226}$Ra is known as decaying with emission of $\alpha$ particle 
to $^{222}$Rn with $T_{1/2} = 1600$ y \cite{ToI98} practically with 100\%
(small probability of $3.2\times10^{-9}$\% exists to emit $^{14}_{~6}$C clusters \cite{ToI98}). 
However, $2\beta$ decay of this
nuclide is also energetically possible with energy release $Q_{2\beta} = 472\pm5$ keV 
\cite{Wan12}. In this case we should see the following chain of subsequent decays:
\begin{equation}\begin{split}
& ^{226}_{~88}\text{Ra}~\xrightarrow[472~\text{keV}]{2\beta?}~
  ^{226}_{~90}\text{Th}~\xrightarrow[6451~\text{keV}]{\alpha~~30.57~\text{m}}~
  ^{222}_{~88}\text{Ra}~\xrightarrow[6679~\text{keV}]{\alpha~~38.0~\text{s}}~ \\
& ^{218}_{~86}\text{Rn}~\xrightarrow[7263~\text{keV}]{\alpha~~35~\text{ms}}~ 
  ^{214}_{~84}\text{Po}~\xrightarrow[7833~\text{keV}]{\alpha~~164.3~\mu\text{s}}~
  ^{210}_{~82}\text{Pb (22.3 y)}.
\end{split}\end{equation}

This chain is very similar to the chain in Eq. (2). Now, instead to look for the 
sequence of decays $^{222}$Fr $\to$ $^{222}$Ra $\to$ $^{218}$Rn considered in section 4.1, 
we should search for $^{226}$Th $\to$ $^{222}$Ra $\to$ $^{218}$Rn sub-chain having initial decay
with energy between 2000 and 2502 keV ($^{226}$Th $E_\alpha$ in the $\gamma$ 
scale of the BaF$_2$ scintillator $\pm$ FWHM$_\alpha$) and mean time characteristic for 
$\alpha$ events (the last two steps are the same as in section 4.1).
The procedure gives 
$\lim S = 5.4\times10^3$ counts and 
$\varepsilon = 0.833$. Using formula (1) with 
$R^\alpha = 13.4$ Bq and 
$T_{1/2}^\alpha = 1600$ yr, one gets:

$$T_{1/2}^{2\beta(0\nu+2\nu)}(^{226}\text{Ra}) > 1.2\times10^6~\text{y at 90\% C.L.}$$

\noindent
or, in terms of the branching ratio, $B < 0.13$\%.
This result is $2-3$ orders of magnitude better than the limits obtained in \cite{Tre05}
($4.1\times10^4$ y for $2\beta0\nu$ and $4.5\times10^3$ y for $2\beta2\nu$ at 68\% C.L.).

\subsection{$2\beta$ decay of $^{212}$Pb}

The situation with $^{212}$Pb is very similar to that considered in section 4.2 for $^{222}$Rn.
While $^{212}$Pb usually transforms to $^{212}$Po through two subsequent single $\beta$ decays:
$$
^{212}_{~82}\text{Pb}~\xrightarrow[574~\text{keV}]{\beta~~10.64~\text{h}}~~
^{212}_{~83}\text{Bi}~\xrightarrow[2254~\text{keV}]{\beta~~60.55~\text{m}}~~
^{212}_{~84}\text{Po (299 ns)},
$$
in principle, it can jump to $^{212}$Po in one step through $2\beta$ decay:
$$
^{212}_{~82}\text{Pb}~\xrightarrow[2828~\text{keV}]{2\beta}~~
^{212}_{~84}\text{Po (299 ns)}.
$$

In this case we can use the energy spectrum of $\beta$ events already obtained by the 
analysis of the fast Bi-Po events (see Fig. 6). The high energy part of the spectrum is shown
in Fig. 9 together with the maximal $^{212}$Pb $2\beta0\nu$ and $2\beta2\nu$ effects 
consistent with the experimental data (the corresponding values of areas are equal to
$\lim S = 46.6$ counts for $2\beta0\nu$ and $1.1\times10^5$ counts for $2\beta2\nu$,
which take into account also the statistical uncertainties in the experimental data). 
Because in chain of decays $^{228}$Th $\to$ ... $\to$ $^{212}$Pb all intermediate nuclides
have half-lives small (from seconds to days) in comparison with time elapsed from
the BaF$_2$ crystal growth (few years), $^{212}$Pb is in secular equilibrium with $^{228}$Th,
and we can use for its activity the value of 
$R^\beta = 2.31$ Bq (the same as for $^{228}$Th, see Table 1).
Using Eq. (1) and 
$\varepsilon = 0.92$,
$T_{1/2}^\beta = 10.64$ h, 
one obtains:

$$T_{1/2}^{2\beta0\nu}(^{212}\text{Pb}) > 20~\text{y at 90\% C.L.,}$$
$$T_{1/2}^{2\beta2\nu}(^{212}\text{Pb}) > 75~\text{h at 90\% C.L.}$$

The value for $2\beta0\nu$ decay is 3 times better than that obtained in \cite{Tre05}
(6.7 y at 68\% C.L.). 
The corresponding limits on the branching ratios are:
$B_{2\beta0\nu} < 6.0\times10^{-3}\%$, $B_{2\beta2\nu} < 14\%$.

~

All the obtained results on $T_{1/2}$'s are summarized in Table 2.

\nopagebreak
\begin{figure}[htb]
\begin{center}
\mbox{\epsfig{figure=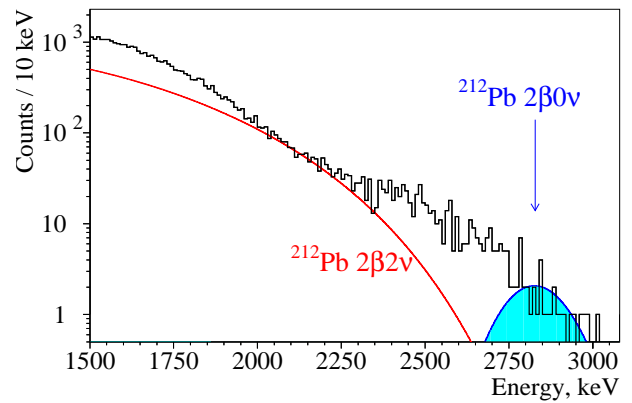,width=8.0cm}}
\caption{(Color online) The energy spectrum of $\beta$ events in the fast decay 
chain $^{212}$Bi--$^{212}$Po together with maximal effects of $^{212}$Pb $2\beta0\nu$ 
and $2\beta2\nu$ decays consistent with the experimental data.}
\end{center}
\end{figure}

\begin{table*}[!h]
\caption{Branching ratios and  half-life values or limits (at 90\% C.L.) obtained in this work
in comparison with other results.
$T_{1/2}$ limits in \cite{Tre05} were given at 68\% C.L.}
\begin{center}
\begin{tabular}{lllllll}
\hline
Nuclide    & \multicolumn{2}{l}{Main channel of}                  & \multicolumn{4}{l}{$T_{1/2}$ (and branching ratio, $B$)}                                                           \\
~          & \multicolumn{2}{l}{decay and $T_{1/2}$ \cite{ToI98}} & \multicolumn{2}{l}{This work} & \multicolumn{2}{l}{Other works}                                             \\
\hline
~ & ~ & ~ & ~ & ~ & ~ \\
$^{212}$Po & $\alpha$ & $299\pm2$ ns                              & \multicolumn{2}{l}{298.8$\pm$0.8(stat.)$\pm$1.4(syst.)}      & 294.7$\pm$0.6(stat.)$\pm$0.8(syst.) & \cite{Bel13} \\
~          & ~        &                                           &                                   &                          & $299\pm2$                           & \cite{Bro05} \\
$^{212}$Pb & $\beta$  & 10.64 h                                   & $2\beta2\nu$~~~$>75$ h            & ($B<14\%$)               & $>146$ h                            & \cite{Tre05} \\
~          & ~        & ~                                         & $2\beta0\nu$~~~$>20$ y            & ($B<6.0\times10^{-3}\%$) & $>6.7$ y                            & \cite{Tre05} \\
$^{222}$Rn & $\alpha$ & 3.8235 d                                  & $\beta$~~~~~~~ $>8.0$ y           & ($B<0.13\%$)             & --                                  & ~            \\
~          & ~        & ~                                         & $2\beta2\nu$~~~$>8.0$ y           & ($B<0.13\%$)             & $>40$ d                             & \cite{Tre05} \\
~          & ~        & ~                                         & $2\beta0\nu$~~~$>8.0$ y           & ($B<0.13\%$)             & $>2.8$ y                            & \cite{Tre05} \\
$^{226}$Ra & $\alpha$ & 1600 y                                    & $2\beta2\nu$~~~$>1.2\times10^6$ y & ($B<0.13\%$)             & $>4.5\times10^3$ y                  & \cite{Tre05} \\
~          & ~        & ~                                         & $2\beta0\nu$~~~$>1.2\times10^6$ y & ($B<0.13\%$)             & $>4.1\times10^4$ y                  & \cite{Tre05} \\
~ & ~ & ~ & ~ & ~ & ~ \\
\hline
\end{tabular}
\end{center}
\end{table*}

\section{Discussion and conclusions}

The radioactive contaminations of a BaF$_2$ crystal scintillator were
estimated to be at level of few Bq/kg for $^{226}$Ra and $^{228}$Th. Taking
into account 3 orders of magnitude lower activity of $^{238}$U and $^{232}$Th
(only limits $<$0.0002 Bq/kg for $^{238}$U and $<$0.004 Bq/kg for $^{232}$Th were
obtained) and broken equilibrium in the chains, one can conclude that the
BaF$_2$ crystal is contaminated by radium (which is chemically close to barium).
The response of the BaF$_2$
crystal scintillator to $\alpha$ particles has been investigated in a wide energy
interval (4.8 -- 9.0 MeV) and the capability of pulse-shape discrimination between $\alpha$ particles and
$\gamma$ quanta (electrons) has been demonstrated.

The analysis of the distribution of the time intervals between $\beta$ and $\alpha$
decays in the fast Bi-Po chains allowed us to estimate the half-life of $^{212}$Po
as $T_{1/2}$ = 298.8$\pm$0.8(stat.) $\pm$1.4(syst.) ns, which is
in agreement with the table value \cite{ToI98,Bro05}.

First limit on $\beta$ decay of $^{222}$Rn was found as 
$T_{1/2}^\beta(^{222}$Rn) $> 8.0$ y at 90\% C.L.
This is still quite far from the theoretical estimation  
$T_{1/2}^\beta(^{222}$Rn$) = 4.8\times10^5$ y (for $Q_\beta=24$ keV).
The half-life limits of $^{212}$Pb, $^{222}$Rn and $^{226}$Ra relatively to $2\beta$
decays were also improved in comparison to the earlier work. 
The big dead time of 1.65 ms in the present measurements did not allow us to use
decay of $^{214}$Po as the last step in searching for specific chains of events.
Therefore, the obtained results can be highly improved with a detector with smaller dead time
and better energy resolution (in comparison with FWHM $\simeq12\%$ at 1 MeV for the used 
BaF$_2$ crystal scintillator).

The contamination of the BaF$_2$ crystal by radium is the
main problem in applications of this scintillator to search for double beta decay
of barium. An R\&D of methods to purify barium from radium traces is in progress
at the Gran Sasso National Laboratories with an aim to develop radiopure
BaF$_2$ crystal scintillators to search for double beta decay of
$^{130}$Ba and $^{132}$Ba. Such a counting experiment is of particular interest, taking into
account positive indications obtained in two geochemical experiments on double beta decay of
$^{130}$Ba.

\end{document}